\documentclass[aps,prb,reprint,amsmath,superscriptaddress,floatfix]{revtex4-2}

\usepackage{multirow}
\usepackage{graphicx}
\usepackage{booktabs}

\begin{document}

\title{Three-dimensional checkerboard spin structure on a breathing pyrochlore lattice}

\author{Margarita G. Dronova}
\affiliation{Okinawa Institute of Science and Technology Graduate University, Onna, Okinawa 904-0495, Japan}
\author{Václav Petříček}
\affiliation{Institute of Physics, Czech Academy of Sciences, Prague, Czech Republic}
\author{Zachary Morgan}
\affiliation{Neutron Scattering Division, Oak Ridge National Laboratory, Oak Ridge, Tennessee 37831, USA}
\author{Feng Ye}
\affiliation{Neutron Scattering Division, Oak Ridge National Laboratory, Oak Ridge, Tennessee 37831, USA}
\author{Daniel M. Silevitch}
\affiliation{Division of Physics, Mathematics, and Astronomy, California Institute of Technology, Pasadena, California 91125, USA}
\author{Yejun Feng}
\email{To whom correspondence should be addressed. E-mail: yejun@oist.jp}
\affiliation{Okinawa Institute of Science and Technology Graduate University, Onna, Okinawa 904-0495, Japan}

\begin{abstract}
The standard approach to realize a spin liquid state is through magnetically frustrated states, relying on ingredients such as the lattice geometry, dimensionality, and magnetic interaction type of the spins. While Heisenberg spins on a pyrochlore lattice with only  antiferromagnetic  nearest-neighbors interactions are theoretically proven disordered, spins in real systems generally include longer-range interactions. The spatial correlations at longer distances typically stabilize a long-range order rather than enhancing a spin liquid state. Both states can, however, be destroyed by short-range static correlations introduced by chemical disorder. Here, using disorder-free specimens with a clear long-range antiferromagnetic order, we refine the spin structure of the Heisenberg spinel ZnFe$_2$O$_4$ through neutron magnetic diffraction. The unique wave vector $(1, 0, 1/2)$ leads to a spin structure that can be viewed as alternatively stacked ferromagnetic and antiferromagnetic tetrahedra in a three-dimensional checkerboard form. Stable coexistence of these opposing types of clusters is enabled by the bipartite breathing-pyrochlore crystal structure, leading to a second order phase transition at 10 K. The diffraction intensity of ZnFe$_2$O$_4$ is an exact complement to the inelastic scattering intensity of several chromate spinel systems which are regarded as model classical spin liquids. Our results challenge this attribution, and suggest instead of the six-spin ring-mode, spin excitations in chromate spinels are closely related to the $(1, 0, 1/2)$ type of spin order and the four-spin ferromagnetic cluster locally at one tetrahedron. 
\end{abstract}

\date{\today}

\maketitle

\section{Introduction}

The three-dimensional pyrochlore structure has been of major research interest since Verwey and Anderson’s original work on charge and magnetic behavior in spinel oxides \cite{Verwey1939,Anderson1956}. If only the nearest neighbor interaction are considered, there is no long-range magnetic order for both ferromagnetic Ising \cite{Anderson1956}  and antiferromagnetic Heisenberg \cite{Moessner1998}  spins, and the geometrical structure would drive the spins into a disordered configuration. Real systems have magnetic interactions beyond nearest-neighbor, which combined with additional anisotropic effects such as crystal fields and Dzyaloshinsky-Moria exchange interactions,  induce many pyrochlore systems to form long-range magnetically ordered states at finite temperatures \cite{Anderson1956,Gardner2010} . For spins residing on a pyrochlore sublattice, most magnetic structures have relatively simple wavevectors such as $(0, 0, 0)$ and $(1/2, 1/2, 1/2)$ \cite{Gardner2010,Yamaura2012,Aroyo2006}.  Here, we consider a unique exception to that trend, the spinel  ZnFe$_2$O$_4$. 

ZnFe$_2$O$_4$ has been studied by both neutron powder \cite{Hastings1953,Hastings1956,Fayek1970,Boucher1970,Konig1970,Schiessl1996,Burghart2000,Usa2004} and single crystal \cite{Kamazawa1999,Kamazawa2003,Sandemann2022} scattering  since the dawn of neutron magnetic scattering seven decades ago. Despite early theoretical interest \cite{Anderson1956}, there is no understanding of its magnetic space group and the antiferromagnetic spin structure. The antiferromagnetic spin structure is known to have wavevector $(1, 0, 1/2)$ below $T_N=9.95$~K, and is strongly suspected to be of a non-collinear form \cite{Fayek1970,Boucher1970,Konig1970}. The wavevector is  unique for spins on a cubic pyrochlore sublattice. While cubic GdInCu$_4$ and HoInCu$_4$ also order magnetically with wavevector $(1, 0, 1/2)$ \cite{Nakamura1999,Stockert2020}, the moments reside on a simple FCC lattice and magnetic interactions are mediated by itinerant electrons. For pyrochlore-structured spin assemblies, this wavevector has only been reported in ZnCr$_2$O$_4$ and MgCr$_2$O$_4$ \cite{Ji2009, Gao2018}. Unfortunately, ZnCr$_2$O$_4$ and MgCr$_2$O$_4$ both have strong first-order lattice distortions to at least the tetragonal symmetry upon  magnetic ordering \cite{Kemei2013}. As the ground state is composed of three or four coexisting spin orders, the $(1, 0, 1/2)$ antiferromagnetic spin structure has not been fully resolved \cite{Ji2009, Gao2018}. Indeed, the listed spin structure in Fig. 4c of Ref. \cite{Gao2018} for MgCr$_2$O$_4$ has no applicable symmetry, and the spin arrangement can potentially violate their  presumption of no net moment per individual tetrahedron; in two tetrahedra not specifically drawn in the figure, the spin clusters appear ferromagnetic. By contrast, ZnFe$_2$O$_4$ provides a clean model antiferromagnet with a single $(1, 0, 1/2)$ vector in the cubic pyrochlore lattice, and the magnetic phase transition is of second-order nature as revealed by heat capacity measurements \cite{Dronova2022}. 

\begin{figure}[tb]
    \centering
    \includegraphics[width=3in]{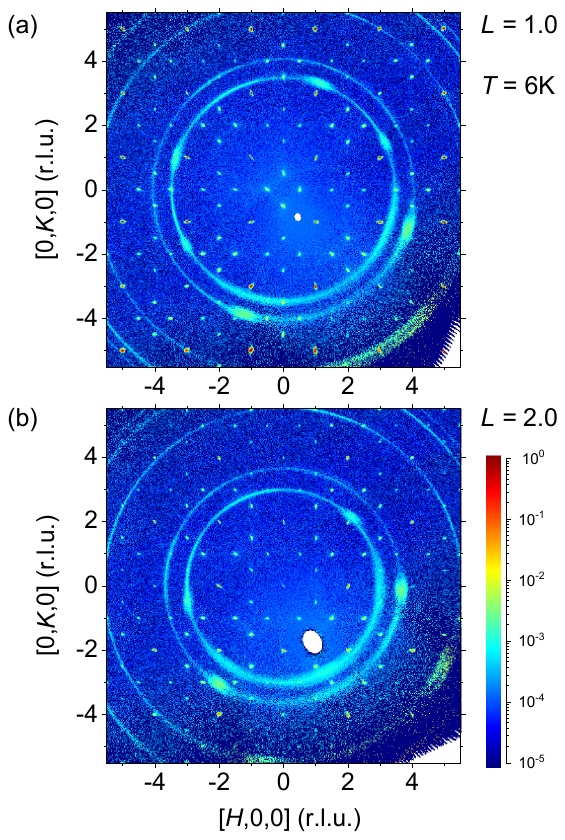}
    \caption{Representative single-crystal diffraction patterns of ZnFe$_2$O$_4$. Unsymmetrized diffraction intensities of both lattice and antiferromagnetism are presented as $H-K$ plane slices at two $L$ values of 1.0 and 2.0. All intensities have been integrated over a thickness of 0.05 r.l.u. along $L$. The data is collected from a piece of crystal in the clean limit \cite{Dronova2022}, and measured below $T_N$. The magnetic diffraction signal, such as $(0, 1/2, 1)$, is resolution limited and shows no sign of diffuse scattering. The presence of $(4n+2, 4m, 0)$ type of reflections indicates the $F\bar{4}3m$ space group instead of $Fd\bar{3}m$. The ring structure is from diffraction of the aluminum sample holder, as no background has been subtracted. For diffraction patterns with $L$ values of 0.0 and 0.5, the readers are referred to Ref. \cite{Dronova2022}. Intensities are in arbitrary units and normalized to the highest count in one pixel. }
    \label{fig:F1}
\end{figure}

Neutron magnetic diffraction studies of ZnFe$_2$O$_4$ in the literature \cite{Hastings1956,Fayek1970,Boucher1970,Konig1970,Schiessl1996,Kamazawa1999,Burghart2000,Kamazawa2003,Usa2004,Sandemann2022} all exhibit a large amount of spin diffuse scattering  originating from disorder \cite{Dronova2022}. It is well documented that mechanical grinding can significantly increase the level of inversion disorder in ZnFe$_2$O$_4$,  greatly affecting its magnetic properties \cite{Usa2004}. Neutron magnetic powder diffraction is also intrinsically incapable of determining a non-collinear  spin structure without an initial presumption \cite{Roth1958,Fayek1970}. Recently, we demonstrated the growth of  ZnFe$_2$O$_4$ single crystals with high stoichiometry and the clean structural limit with a minimal amount of inversion disorder \cite{Dronova2022}. The long-range antiferromagnetic order was clearly established by neutron single crystal diffraction with no signature of spin diffuse scattering (Fig. \ref{fig:F1}). Single crystal magnetic diffraction lifts the $d$-spacing degeneracy between reflections, and the enhanced signal-to-background ratio allows for the extraction of weak magnetic reflections to very high transferred momentum. However, the improved quality of single crystals leads to severe extinction effects, reducing the diffraction intensity to as low as 5\% \cite{Lawrence1973,Dronova2022}. For single crystal diffraction measurements using a monochromatized beam, the severity of the extinction effect is often indirectly inferred by unrealistic values of refined thermal parameters \cite{Sequeira1972}. With the advancement of chopping schemes and detector technology in recent years, neutron elastic scattering can be measured with the full information of wavelength $\lambda$ and diffraction $2\theta$ \cite{Jogl2011,Ye2018}, and the extinction effect can be directly visualized as a continuous function of $\lambda$ \cite{Dronova2022} (Appendix~\ref{app:extinct}). Such an experimental design provides an opportunity to have the extinction effect examined in every reflection family, which is an advantage beyond simply  improving statistics by incorporating a large number of diffraction events. As continuous wavelength-based neutron diffraction at spallation neutron sources has become routine over the last decade \cite{Zikovsky2011,Jogl2011,Schultz2014,Ye2018}, it is now both imperative and opportunistic to quantitatively take the extinction effect into account for better refinement. The improving understanding of magnetic space groups in the last two decades has also been a key enabler for performing magnetic structure analyses in single crystals\cite{Aroyo2006,Petricek2010,Petricek2023}. 

Here we report the spin structure of ZnFe$_2$O$_4$ through single crystal diffraction and refinement at the continuous-wavelength neutron scattering beamlines CORELLI and TOPAZ. With high-fidelity extraction of the diffraction intensity, correction of the extinction effect over the broad wavelength range, and analysis of the symmetry conditions, we deduce the antiferromagnetic spin arrangement in the parent crystal space group $F\bar{4}3m$ of broken inversion symmetry, and identify it as based on representation $mW1$ in the magnetic space group $I_c\bar{4}2m$. Surprisingly, this unique spin structure has spins align ferromagnetically and antiferromagnetically on corner connected tetrahedra, forming a three-dimensional checkerboard pattern. This checkerboard pattern becomes possible because of the $F\bar{4}3m$ space group, as the breathing pyrochlore lattice accommodates both FM and AF spin clusters on two different sized tetrahedra. Our revelation of a single $(1, 0, 1/2)$ ordered antiferromagnetic state on a pyrochlore lattice has profound implications to the existing understanding of the classical spin liquid systems ZnCr$_2$O$_4$ and MgCr$_2$O$_4$. 

\section{Experimental Methods}

\subsection{Single crystal preparation}
Single crystals of ZnFe$_2$O$_4$  were grown from high-temperature molten solution of anhydrous borax \cite{Dronova2022}. The proportions of ZnO and Fe$_2$O$_3$ powders in the initial solution were finely tuned to make the molar ratio of Fe and Zn ions in the grown crystals stoichiometric (within $\pm0.005$ of 2.000). Furthermore, by growing from different starting temperature, the level of inversion disorder of spinel oxide can be controlled \cite{Dronova2022}.

\subsection{Phase diagram characterization}
The field-temperature $H$-$T$ magnetic phase diagram of ZnFe$_2$O$_4$ has been characterized by both heat capacity and magnetic susceptibility measurements. Three pieces of stoichiometric crystals in the clean limit were used; all were grown from 850 $^\circ$C to 750 $^\circ$C \cite{Dronova2022}, including the one in Supplementary Fig. 1d of Ref. \cite{Dronova2022}. These samples have a Curie-Weiss temperature $T_{CW}=-25$~K \cite{Dronova2022}. The magnetization  was measured using the vibrating sample magnetometer (VSM) of a Physical Property Measurement System (PPMS) (Dynacool-14T, Quantum Design, Inc.), with temperature varying from 2 K to 400 K and field up to 14 Tesla, applied on two pieces of crystal along the (1, 1, 0) and (1, 0, 0) directions. Heat capacity was measured on two crystals with field direction applied along the $(1, 1, 1)$ and $(1, 0, 0)$ directions respectively in a Dynacool-9T PPMS. 

\subsection{Wavelength-resolved neutron diffraction}
Structural and magnetic characterizations of stoichiometric ZnFe$_2$O$_4$ were carried out at two neutron elastic scattering beamlines: TOPAZ (BL-12) and CORELLI (BL-9) at the Spallation Neutron Source of Oak Ridge National Laboratory \cite{Zikovsky2011, Jogl2011, Schultz2014, Sullivan2018, Ye2018}. Both beamlines utilize unpolarized incident neutrons of continuous wavelength over slightly different ranges, from 0.6 to 3.5 Å for TOPAZ and from 0.8 to 2.4 Å for CORELLI. The short wavelength cutoff is set by the focusing limit of the high-flux beam guide, a key component that enables probing sub-mm sized samples. The long wavelength limit is set by the SNS pulse frequency (60 Hz) and the distance between the neutron moderator and the sample position at the beamline \cite{Schultz2014, Ye2018}. 

Most of our neutron data were collected from one piece of stoichiometric single crystal that was measured at TOPAZ at 100 K and at CORELLI at 6 K. The single crystal was grown from 1000 $^\circ$C to 850 $^\circ$C, weighs 7.4 mg, and is  1.2 mm in diameter (Supplementary Fig. 1c of Ref. \cite{Dronova2022}). Its magnetic susceptibility was reported earlier \cite{Dronova2022} with a $T_{CW}=-20$ K. Although CORELLI is specialized for studying elastic diffuse scattering, the 1000 $^\circ$C-grown single crystal does not demonstrate magnetic diffuse scattering in the ordered phase but instead exhibits a pattern of sharp diffraction spots from both antiferromagnetism and the lattice (Fig. \ref{fig:F1}, and Ref. \cite{Dronova2022}). The comparison in Fig. \ref{fig:F8} uses magnetic diffuse scattering data taken at CORELLI from a mosaic assembly of stoichiometric crystals (Supplementary Fig. 1e of Ref. \cite{Dronova2022}); these crystals were grown at temperatures from 1250 $^\circ$C to 950 $^\circ$C, resuliting in a high level of inversion disorder. 

Despite their similarity of continuous wavelength neutron diffraction, data collections at TOPAZ and CORELLI have different emphases on the coverage of reciprocal space. At both beamlines, single crystal samples are placed at a series of fixed angular positions for a certain amount of time while diffraction happens for neutrons with wavelengths that satisfy Bragg’s law. TOPAZ covers a large 2$\theta$ range from 15$^\circ$ to 155$^\circ$, allowing the same reflection to be measured over a large wavelength range. At TOPAZ, 35 frames were collected for the current sample over a total flux of 94.3 C of proton charge, positioned at largely different $\omega,\chi, \phi$ angular positions. The choices of crystal angular placements emphasized the coverage of many families of reflections in  reciprocal space but not necessarily continuous \cite{Zikovsky2011}. On the other hand, CORELLI provides a continuous coverage of the reciprocal space in order to reconstruct the diffuse scattering pattern without symmetrization. Typically, the crystal is rotated only along the vertical axis, and in our experiment by 1.5$^\circ$ steps over the whole 360$^\circ$ range for a total of 240 frames. The flux of 0.32 C of proton charge per frame we used at CORELLI is much shorter than that at TOPAZ. Additional details on the data reduction process are given in Appendix~\ref{app:reduce} and a discussion of the determination of wavelength-dependent extinction effects is provided in Appendix~\ref{app:extinct}.

\section{Results: Lattice structure refinement}

\begin{figure}[!tb]
    \centering
    \includegraphics[width=3.5in]{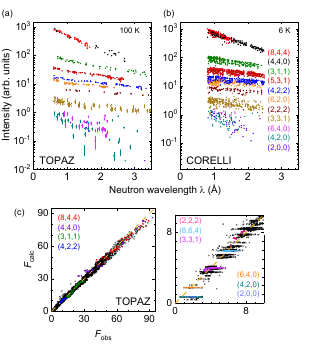}
    \caption{Single crystal diffraction intensities at (a) TOPAZ and (b) CORELLI, color grouped by reflection families and plotted as a function of measurement wavelength. For strong reflection families, the slanted wavelength dependence is a signature of the extinction effect. However, weak reflections also demonstrate a noticeable wavelength dependence that is not expected. This remaining issue in data reduction will be improved in future work.  (c) Calculated versus observed lattice structure factors $F_{calc}$ vs. $F_{obs}$ of all data (black crosses) taken at TOPAZ; zoomed-in view shows the comparison for reflections of weak intensity. Colored points mark selected families as listed in the legend. While the ideal behavior is the dashed diagonal line (orange), for strong reflections, the bending of the curve indicates remaining incompletely corrected extinction effect. }
    \label{fig:F3}
\end{figure}

Representative CORELLI single-crystal diffraction patterns for ZnFe$_2$O$_4$ are shown in Fig. \ref{fig:F1}, and  intensities of several reflection families and their wavelength dependence are plotted in Fig. \ref{fig:F3} for measurements of the same sample at CORELLI and TOPAZ. For strong reflections such as $(8, 4, 0)$ and $(4, 4, 0)$, the wavelength dependence in Fig. \ref{fig:F3} clearly demonstrates a severe extinction effect; for $(4, 4, 0)$,  $I(2.7\text{\AA})/I(0.6\text{\AA})\approx0.07$. Furthermore, intensity ratios between the strongest reflections such as $(8, 4, 4)$ and $(4, 4, 0)$ and the weakest reflections such as $(2, 0, 0)$ and $(4, 2, 0)$ are more than $10^3$. Overall, we have over four decades of dynamic range in measured signal strength (Appendix~\ref{app:extinct}). Previously, in Ref. \cite{Dronova2022}, the data reduction procedure produced much less dynamic range, with the largest intensity ratio being less than 200, indicating that reductions of past single crystal diffraction studies at TOPAZ could have been yielded overestimated intensities of weak reflections. 

The space group of ZnFe$_2$O$_4$ was recently determined to be $F\bar{4}3m$ \cite{Dronova2022} rather than $Fd\bar{3}m$ as reported in the literature \cite{Boucher1970,Konig1970,Sandemann2023}. Our raw data is reduced using general selection rules for a face-centered Bravais lattice with $h,k,l$ either all even or all odd; all reflections allowed in the $F\bar{4}3m$ space group but forbidden in $Fd\bar{3}m$ such as $(0, 0, 2)$ and $(4, 2, 0)$ are collected with non-vanishing intensities (Fig. \ref{fig:F3}). Using a data reduction based on the selection rules of the primary Bravais lattice does not introduce forbidden reflections of the $F\bar{4}3m$ space group, such as  mixed even-odd indices like $(0, 0, 1)$ and $(0, 1, 1)$. This can be directly visualized from the unsymmetrized data plotted in Fig. \ref{fig:F1}. Multiple scattering has been argued in Ref. \cite{Sandemann2023} to explain our observation of forbidden peaks of  $Fd\bar{3}m$. As extensively discussed in Ref. \cite{Dronova2022}, multiple scattering and other potential artifacts such as harmonics can be ruled out utilizing the simultaneously recorded neutron wavelength $\lambda$ and diffraction angle $2\theta$ provided by the continuous wavelength diffraction technique \cite{Ye2018}. Furthermore, both techniques of Ref. \cite{Sandemann2023}, single crystal azimuthal measurements using a lab-based x-ray tube source and powder x-ray diffraction, can not provide the necessary dynamic range to rule out  the presence of weak peaks. It is well understood that the presence of strong multiple scattering does not rule out the presence of weak  real signals \cite{Wang2020a}, as the technical challenge always resides at the sensitivity of low intensities when the multiple scattering becomes weak. 

In Ref. \cite{Dronova2022}, the extinction effect was corrected for each individual family to extract $F_{hkl}^2$ in the $\lambda=0$ limit; approximately 25\% of the strong reflection families with large extinction were removed from the refinement. The manual selection process \cite{Sequeira1972,Dronova2022} was effective but ad hoc, introducing the ingredient of discretion.  Here, we seek a global refinement of all diffraction events, with the wavelength and structure-factor dependent extinction effect fully incorporated in the refinement software (Appendix~\ref{app:extinct}). 

The refined lattice positions and thermal parameters from both TOPAZ and CORELLI measurements are summarized in Table~\ref{app:Table1}. The two lattice structures at 100 K and 6 K, across the antiferromagnetic phase boundary, are both consistent with $F\bar{4}3m$, indicating a second order antiferromagnetic phase transition. The amount of inversion symmetry breaking appears to increase going from TOPAZ data at 100K to CORELLI data at 6 K, potentially suggesting a magnetostrictive nature. At 6 K, the two oxygen positions in the current refinement have a larger amount of inversion symmetry breaking than the values listed in Ref. \cite{Dronova2022}, while the iron position  has a smaller deviation from the inversion symmetry position than that given in Ref. \cite{Dronova2022}. We attribute the difference between the current and previous studies to the improved neutron data reduction for intensity integration (Appendix~\ref{app:reduce}) \cite{Morgan}. 

The overall quality of refinement is evaluated by plotting $F_{calc}$ vs. $F_{obs}$ in Fig. \ref{fig:F3}c for all events. Ideally, $F_{calc}$ vs. $F_{obs}$ should follow a  diagonal line if the extinction effect is fully corrected. Here, we observe a slight curvature in the plot, indicating the extinction effect has been corrected but not to the full extent. Several reflection families of high, medium, and low intensities are marked in Fig. \ref{fig:F3}c. For reflection families with strong intensities, such as $(8, 4, 4)$ and $(4, 4, 0)$, there is insufficient correction to the extinction effect. Families of intermediate intensities, such as $(3, 1, 1)$ and $(2, 2, 2)$, $F_{calc}$ vs. $F_{obs}$ follows the diagonal line. 

It would be desirable to use the large statistics of our data to further analyze the inversion disorder in ZnFe$_2$O$_4$ \cite{Dronova2022}. However, both $R$ and $GOF$ values in the current refinement (Table~\ref{app:Table1}) remain relatively large, which indicates the need to fundamentally improve the extinction correction process (Appendix~\ref{app:extinct}). At the current stage, we refrain from further analysis of the inversion disorder. 

\section{Results: Non-collinear antiferromagnetic structure}

\begin{figure}[!tb]
    \centering
    \includegraphics[width=3in]{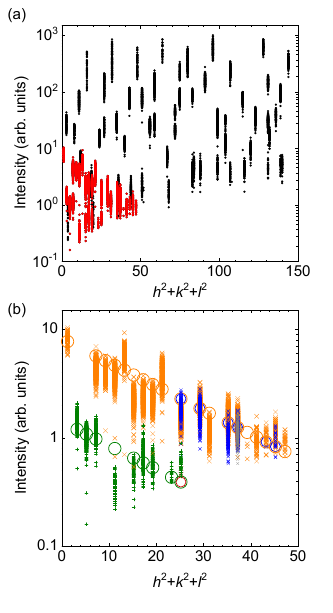}
    \caption{(a) Diffraction intensities of both lattice (black) and magnetism (red), plotted for all events of the sample measured at CORELLI. The intensities are plotted versus $h^2+k^2+l^2\propto q_{hkl}^2$ . Each individual reflection family is contained  within a vertical column; within a same column, there can be a degeneracy of several reflection families. (b) Intensities versus $h^2+k^2+l^2$ for  magnetic diffraction events (crosses). Various colors (orange, green, blue, burgundy, grey) are used to separate different reflection families with degenerate $d$-spacing. A simple functional form $I_M \sim [1+  \cos\left(\pi h/2\right)\cos\left({\pi l/2}\right)] e^{-q^2}$ (circles) is plotted for comparison, with $h$ and $l$ the even and half-integer indices of the reflection $(h, k, l)$. Matching colors are used to relate calculated values to diffraction data of the same reflection family. Concentric circles are used to differentiate the $d$-spacing degeneracy. This functional form captures the nearly binary feature of the diffraction intensities. 
    }
    \label{fig:F4}
\end{figure}

The long-range antiferromagnetic order in ZnFe$_2$O$_4$ was previously explored by  neutron powder diffraction studies in the early 1970s \cite{Fayek1970, Konig1970, Boucher1970};  the focus of more recent neutron magnetic scattering has shifted to disorder-dominated short-range spin correlations \cite{Schiessl1996, Kamazawa1999, Burghart2000, Kamazawa2003, Usa2004, Sandemann2022}. The three powder neutron diffraction studies in Ref. \cite{Fayek1970,Konig1970,Boucher1970}  each collected only 8-13 magnetic peaks, including many degeneracies with reflections of identical $q$. These datasets are inadequate to refine a non-collinear spin structure \cite{Roth1958}  and were instead limited to collinear types of antiferromagnetism in ZnFe$_2$O$_4$; one proposed structure set collinear spins at an angle of $35^\circ-45^\circ$ to the $c$-axis \cite{Fayek1970}. In contrast to powder magnetic diffraction, our single crystal neutron magnetic diffraction from CORELLI collected a total of 7,921 magnetic events across a range of wavelengths, which represent 899 independent reflections and 36 distinctive magnetic reflection families (Fig.~\ref{fig:F4}, Table~\ref{app:Table2}). This more comprehensive dataset allows for refinement of a non-colinear structure, as detailed below. 

\subsection{Maximal point symmetry of the magnetic unit cell}

\begin{figure}[!tb]
    \centering
    \includegraphics[width=2.5in]{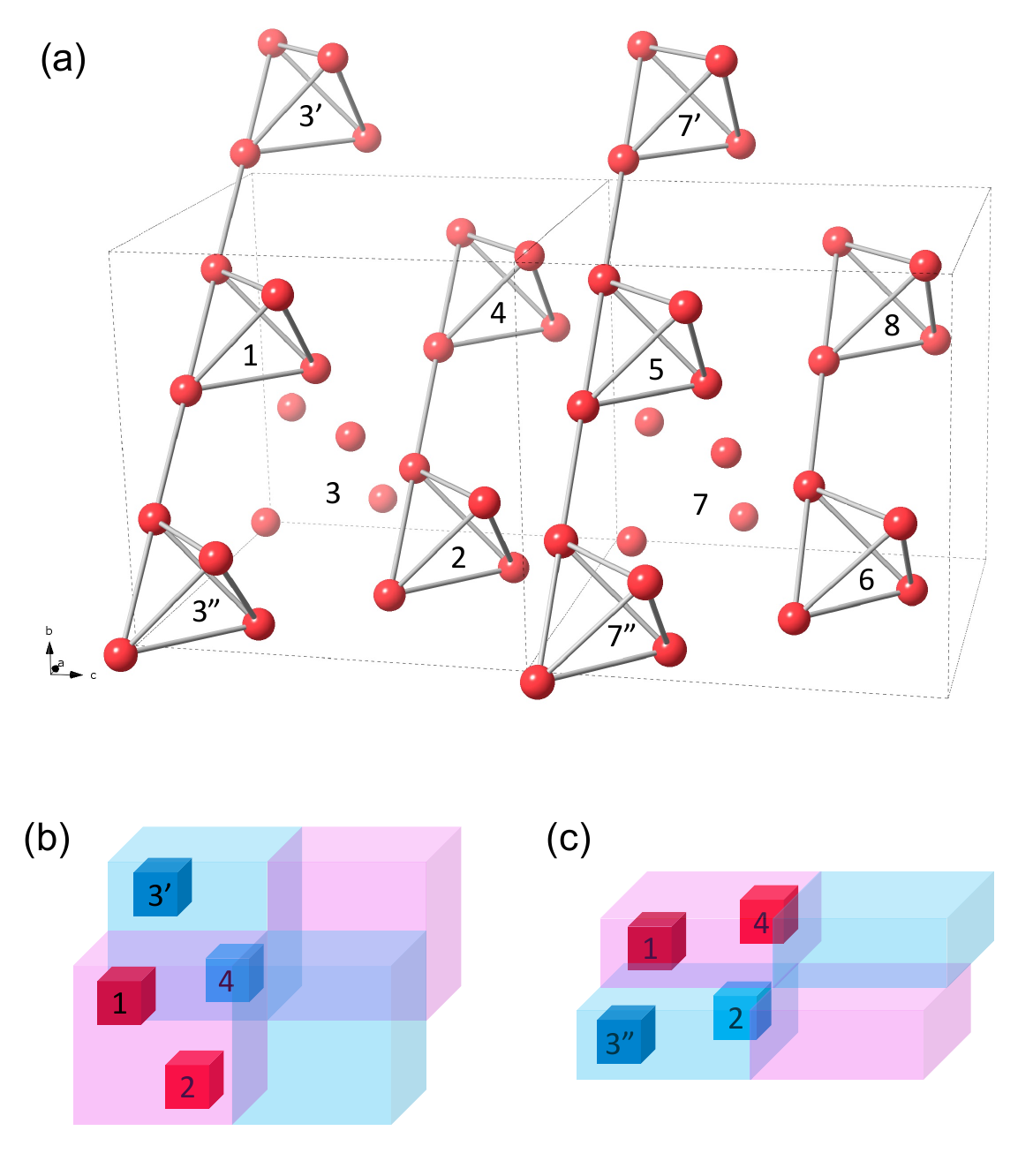}
    \caption{(a) All 32 B-sites of  AB$_2$O$_4$ spinels in a magnetic unit cell of $a\times a\times 2a$ can be grouped into eight tetrahedra marked as 1 to 8. Equivalent tetrahedra in neighboring magnetic unit cells and the inter-tetrahedra bonds are included to demonstrate the translational relationship according to the vector $(a/2, a/2, 0)$. (b-c) The eight tetrahedra can be grouped into four equivalent rectangular blocks (blue and pink) of two tetrahedra each. As the blocks are connected by three vectors $(a/2, a/2, 0)$, $(0, 0, a)$ and $(a/2, a/2, a)$, spins in pink and blue blocks are antiparallel to satisfy the antiferromagnetic wave vector. The equivalency of $(1, 0, 0.5)$ and $(0, 1, 0.5)$ vectors can be viewed as different constructions of blocks in (b) and (c) respectively. }
    \label{fig:F5}
\end{figure}

The basic magnetic wave vector $Q = (1, 0, 1/2)$ (Fig. \ref{fig:F1}), expressed in reciprocal space units of the FCC lattice, doubles the magnetic unit cell of ZnFe$_2$O$_4$ to $a\times a\times 2a$. The wave vector $(1, 0, 1/2)$ is located at the high symmetry point $W$ on the first Brillouin zone boundary of the FCC lattice, where $W$ possesses the $\bar{4}2m$ point group symmetry. For the 32 Fe$^{3+}$ spins in the magnetic unit cell, the structure of $Q$ separates them into four blocks of spins with parallel or antiparallel relationships, connected by translation vectors of $(a/2, a/2, 0)$, $(0, 0, a)$ and $(a/2, a/2, a)$ (Fig. \ref{fig:F5}) \cite{Fayek1970,Konig1970}. With the parent pyrochlore lattice, the magnetic unit cell can be simultaneously inspected through both $(1, 0, 0.5)$ and $(0, 1, 0.5)$ wave vectors (Fig. \ref{fig:F5}), and the spin structure is isotropic within the $a$-$b$ plane. This construction in Fig. \ref{fig:F5} indicates that time inversion $1^\prime$ is a good symmetry operation when it is combined with spatial translations, and the ionic lattice  allows the \textit{nonmagnetic} (\textit{grey}) type of magnetic point group. 

This doubling of the magnetic unit cell assumes a single-$Q$ spin structure. With the presence of three independent wavevectors $(1, 0, 1/2)$, $(0, 1/2, 1)$, and $(1/2, 1, 0)$ and the cubic parent lattice, the antiferromagnetic domains in ZnFe$_2$O$_4$ could be either single- or multi-$Q$ type. We first discuss  spin structures of the single-$Q$ type, treating the multi-$Q$ scenario below. All three single-$Q$ domains have separated diffraction patterns, and within the cubic parent structure, the only type of twinning is the inversion symmetry breaking which does not affect the diffraction intensity \cite{Petricek2010} (Appendix~\ref{app:extinct}).

\subsection{Choices of the magnetic space group }

As the magnetic unit cell doubles the lattice unit cell, the Bravais lattice changes from  face-centered cubic to body-centered tetragonal. A reduction of symmetry in general introduces extra degree of freedom or uncertainties to specify the lattice space group of the magnetic unit cell. Here, the loss of the translational vector $(a/2, 0, a/2)$, the face-center vector in the parent cubic structure, leads to possibilities of both symmorphic ($I\bar{4}2m$, \#121) and non-symmorphic ($I\bar{4}2d$, \#122) types of body-centered tetragonal space groups  to describe the ionic lattice of ZnFe$_2$O$_4$. The difference is that the cubic face-center vector $(a/2, 0, a/2)$ is reincorporated as the gliding and screw axis vector $(a/2, 0, 3a/2)$ in $I\bar{4}2d$. 

From these lattice space groups $I\bar{4}2m$ and $I\bar{4}2d$, only two magnetic space groups of the black-white type of the second kind, $I_c\bar{4}2m$ (\#121.332) and $I_c\bar{4}2d$ (\#122.338), are compatible with the specific magnetic wavevector $(1, 0, 1/2)$ in ZnFe$_2$O$_4$. Both magnetic space groups include lattice space group operations in combination with anti-translations made of $(a/2, a/2, 0)$ or $(0, 0, a)$ and the time reversal operator $1^\prime$. Together with normal operations at the origin $(0, 0, 0)$ and body-center translation vector $(a/2, a/2, a)$, spins in these space groups acquire the parallel/antiparallel relationship as Fig. \ref{fig:F5}.  

The combination of eight $\bar{4}2m$ point group operations, the body-center vector $(a/2, a/2, a)$, and the time reversal operation $1^\prime$ symmetrically connects spins in the magnetic unit cell. It leaves only one independent spin in the space group $I_c\bar{4}2d$ to describe all 32 spins. However, in the space group $I_c\bar{4}2m$, the lack of the gliding vector $(a/2, 0, 3a/2)$ leads to a degeneracy of two symmetry operations on each site under all scenarios, and there are always two independent sets of spins (colored vs. grey  in Fig. \ref{fig:F6}) in the magnetic unit cell. This degeneracy of symmetry operations leads to strong constraints for the spin direction as we discuss below.  

We further explore potential magnetic space groups of lower symmetry. While disordered ZnFe$_2$O$_4$ often demonstrates a ferromagnetic state \cite{Burghart2000}, our clean samples do not have any perceivable FM behavior. We thus rule out magnetic space groups with ferromagnetic type of point groups based on Dzyaloshinsky’s principle \cite{Dzyaloshinsky1958}. We note that Ref. \cite{Boucher1970}  suggested $I_c\bar{4}2d$ as the choice of the highest symmetry magnetic space group but ruled out $I_c\bar{4}2m$ based on molecular field considerations in conjunction with Mossbauer studies. Here, we work under the simple assumption of a body-centered, purely antiferromagnetic state. A survey of the symmetry operation reveals all suitable magnetic space groups are \textit{nonmagnetic} type, including choices in $I_c\bar{4}$ (\#82.42), which is suggested by \textit{JANA}2020 and was implied by Ref. \cite{Konig1970}, and $I_c222$ (\#23.52), which was suggested by Ref. \cite{Boucher1970}. Together with $I_c \bar{4}2m$ and $I_c \bar{4}2d$, these four magnetic space groups are the only choices for antiferromagnetism in ZnFe$_2$O$_4$ with the symmetry no lower than orthorhombic. 

$I_c \bar{4}$ and  $I_c 222$ are both symmorphic space groups and both point groups $\bar{4}$ and $222$ have half of the symmetry operations removed from $\bar{4}2m$, including two mirror planes. The reduced total number of 16 symmetry operations from the point group, BCC-translation, and time reversal $1^\prime$ leaves both magnetic space groups $I_c \bar{4}$ and $I_c 222$ with two independent sets of spins for the whole magnetic unit cell, and with no degeneracy of symmetry operations on each site. 

\subsection{Refinement based on representation analysis}

The preceding analysis can be formally derived through  representation analysis \cite{Bertaut1968}, which is incorporated in \textit{JANA}2020 software \cite{Petricek2010,Petricek2023}. In the paramagnetic phase of ZnFe$_2$O$_4$, the $F\bar{4}3m$ space group of the parent lattice has 24 symmetry operations separated into five classes. With the non-zero wave vector $(1, 0, 1/2)$, a purely rotational point group $\bar{4}$ is derived from the original point group to form the wavevector group $G_0$, which leaves the wavevector invariant \cite{Bertaut1968}. Here, $\bar{4}$ represents the kernel symmetry of the antiferromagnetic $(1, 0, 1/2)$ order, and the point group $\bar{4}2m$ represents the higher level epikernel symmetry because of the mirror planes. We  can thus separate all four choice magnetic space groups into three levels of symmetry, with $I_c \bar{4}2d$ and $I_c \bar{4}2m$ at the epikernel level, $I_c \bar{4}$ at the kernel level, and $I_c 222$, which cannot be directly derived through the epikernel/kernel relationship; $I_c 222$ can only be derived from $I_c \bar{4}2m$ \cite{Boucher1970} based on Landau’s theory of continuous phase transition, but not from $I_c \bar{4}2d$ and $I_c \bar{4}$. 

\begin{figure*}[!tbh]
    \centering
    \includegraphics[width=5.5in,trim = {0.5cm 2cm 1cm 1.5cm}, clip]{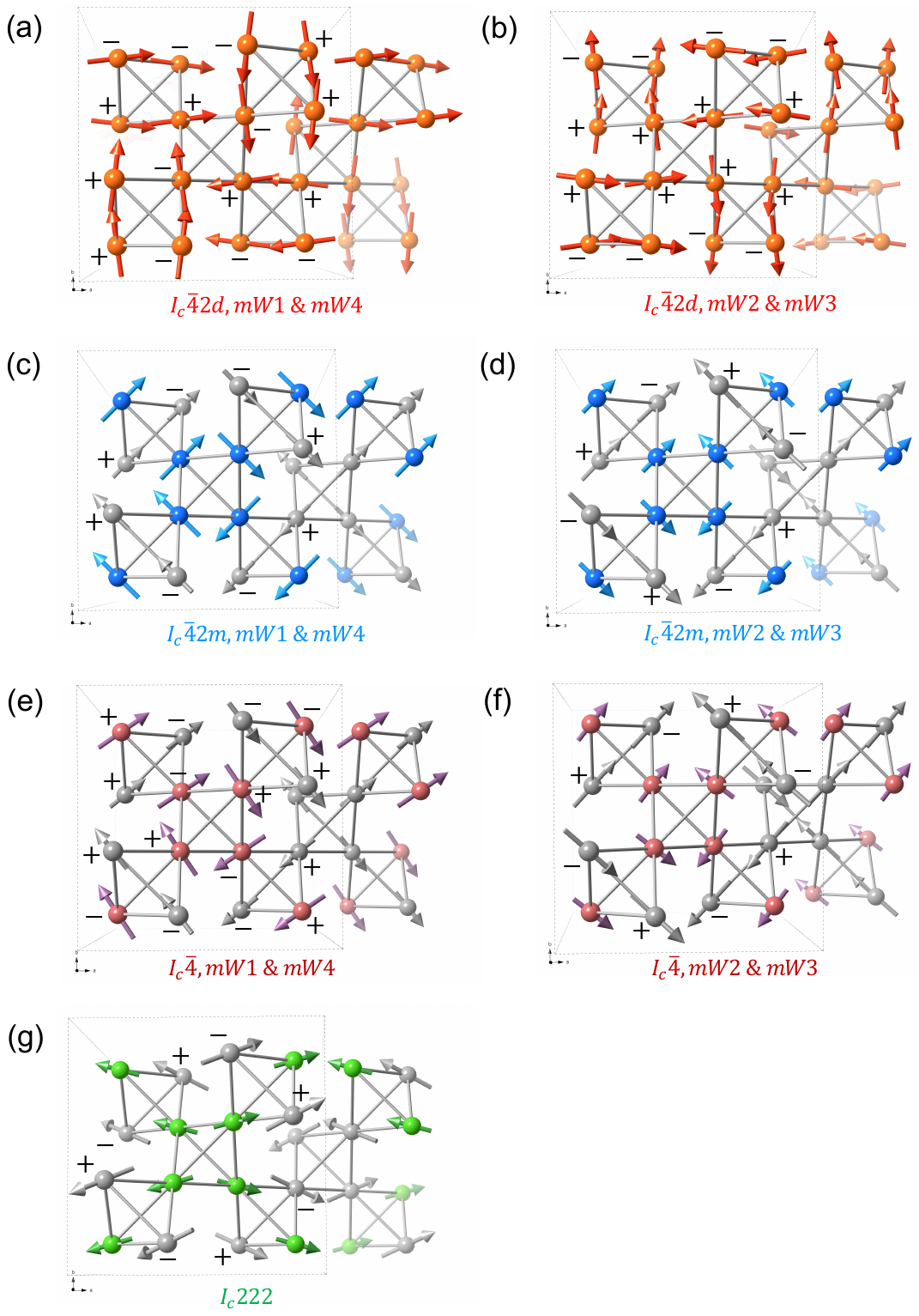}
    \caption{A summary of refined non-collinear spin structure in four different magnetic space groups and their irreps. (a-b) $I_c \bar{4}2d$, with one independent set of spins (red). (c-d)  $I_c \bar{4}2m$, with two independent sets of spins (blue and grey). Spin components in the $a$-$b$ plane are strictly along the $(1, 1, 0)$ or $(1, -1, 0)$ directions for both species. (d-e)  $I_c\bar{4}$, with two independent sets of spins (burgundy and grey). (f) $I_c222$, with two independent sets of spins (green and grey). The schematics are viewed along the $c$-axis and include half of the magnetic cell within the drawn boundary $a\times a$  of the unit cell in the $a$-$b$ plane.  16 spins within half of a magnetic unit cell form four FM clusters and in the middle, there is an AF tetrahedron. Spins in the other half of the unit cell all reverse directions. Moving from the center of the AF tetrahedron by the FCC vector $(a/2, 0, a/2)$, there is a second AF tetrahedron, which is likewise surrounded by FM tetrahedra; only three out of four were drawn here. The perspective has atoms fading in color going towards the opposite $c$-axis direction into the page. $+$ or $-$ markers by each spin indicate positive or negative $c$-axis spin components respectively. For those unmarked, the $c$-axis component is either restricted to zero by symmetry (c, d) or refined to values of statistical zero (f, g). While spin structures refined from irreps $mW1$ and $mW2$ are plotted here and listed in Table~\ref{app:Table2}, other degenerate irreps are also listed under each schematic.}
    \label{fig:F6}
\end{figure*}

As an Abelian group of four elements, $\bar{4}$ has four one-dimensional irreducible representations (irreps); Ref. \cite{Konig1970} listed the irreps together with associated basis functions for ZnFe$_2$O$_4$. Each of the four irreps ($mW1$-$mW4$) can be associated with the three kernel/epikernel space groups we discussed above, based on a different origin of symmetry operations. Using \textit{JANA}2020, we refine spin structures with for all distinct combinations of magnetic space group and irrep. 

For magnetic diffraction data taken from the clean-limit crystal at CORELLI at 6 K, the lattice structure is first refined including the extinction effect, generating the results in Table~\ref{app:Table1}. The magnetic structure is subsequently refined using the updated lattice information. All parameters of atomic positions and thermal parameters are fixed during this stage. This is mainly because for diffraction data at 6 K, all refined lattice thermal parameters $U_{iso}$ are near zero. If they are allowed to float during the refinement of magnetism, the refined values could fluctuate to negative values which would appear unphysical. In the second stage of magnetism refinement, we constrain the extinction effect for both lattice and magnetism to be the same and refine it once again.

In addition to kernel and epikernel space groups, we also include $I_c 222$ as a low-symmetry alternative. The inclusion of space groups of different (and lower) symmetries provides an understanding of certain symmetry elements’ stability. Several irreps have essentially identical refined spin structures in the same magnetic space group, and we can identify seven unique spin structures that are summarized in Fig. \ref{fig:F6} and  Table~\ref{app:Table2}. All  refinements generate similar $R$ and $GOF$ values; these are dominated by systematic issues of extinction correction at the  lattice refinement stage discussed above. 

\subsection{Single-$Q$ vs. multi-$Q$ spin structures}

Before evaluating results under the single-$Q$ scenario, we first explore whether ZnFe$_2$O$_4$ could also be in a multi-$Q$ state. For a triple-$Q$ state, the magnetic unit cell doubles along all three dimensions to $2a\times2a\times2a$. As two spins separated by $(a, 0, 0)$ cannot simultaneously satisfy the parallel and antiparallel conditions set by the $(1, 0, 0.5)$ and $(0.5, 1, 0)$ orders respectively, each $Q$-order in the triple-$Q$ state must come from separated spin components and there will be no combined reflection such as $(1.5,1.5,1.5)$; the diffraction pattern of a multi-$Q$ state would be identical to that of single-$Q$ \cite{Moncton1977}. 

A single-$Q$ state can be unambiguously identified if there is a reduction of lattice symmetry associated with the magnetic phase transition \cite{Rossat-Mignod1987,Feng2013}. However, the antiferromagnetic transition in ZnFe$_2$O$_4$ is of second order nature and demonstrates no observable structure modification. On the other hand, there exist indirect evidence against a multi-$Q$ state, which results from higher order spin-spin interaction terms in the Hamiltonian, and is more often seen in $4f$/$5f$ magnets such as Gd$_2$Ti$_2$O$_7$, CeB$_6$, UO$_2$, and UP \cite{Stewart2004, Rossat-Mignod1987}. The higher order terms could have stronger influence at lower $T$, so multi-$Q$ states emerge at a second phase transition below $T_N$ in Gd$_2$Ti$_2$O$_7$ and UP. As $3d$ spins are of Heisenberg type with little anisotropy in the exchange interaction, a single-Q state is more likely to happen than a multi-$Q$ state \cite{Rossat-Mignod1987}.

\begin{figure}[tbh]
    \centering
    \includegraphics[width=3in]{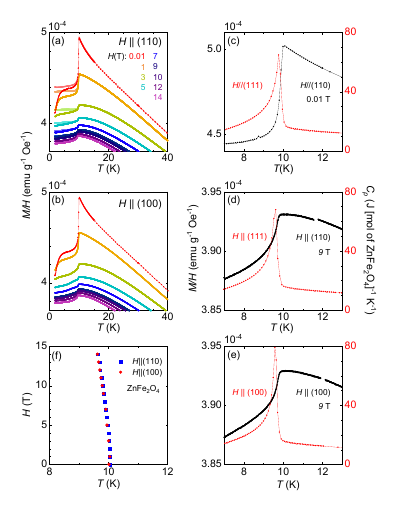}
    \caption{Representative scans of magnetic susceptibility $M/H$ are plotted for (a) $H || (1, 1, 0)$, and (b) $H || (1, 0, 0)$. All curves are  zero-field cooled, except for the lighter traces in panel (b) which are field-cooled . (c, d, e) Heat capacity $C_p$ in the magnetic transition region, with field aligned along (c-d) $(1, 1, 1)$ and (e) $(1, 0, 0)$ axes, are plotted together with magnetic susceptibility at the same field. Both measurements consistently mark the antiferromagnetic phase transition under field. Further measurements of $C_p$ at field intervals of 1-2 T along these two axes (not shown) also reveal no extra phase below $T_N$ to 9 T. (f)  $H$-$T$ phase diagram to 14 T based on magnetic susceptibility data, demonstrating a very weak field dependence and isotropy along all directions.}
    \label{fig:F2}
\end{figure}

A multi-$Q$ state can be verified through anisotropic external fields such as uniaxial pressure or magnetic field; for example, cubic CeB$_6$ transitions from a double-$Q$ to a single-$Q$ state under increasing magnetic field \cite{Rossat-Mignod1987}. We characterized the  antiferromagnetic field-temperature phase diagram of ZnFe$_2$O$_4$ using both  DC magnetic susceptibility and heat capacity. With $H || (1, 0, 0)$ and $H || (1, 1, 0)$, $M/H$ vs. $T$ in Fig.~\ref{fig:F2} has a very weak field dependence. With $H || (1, 0, 0)$ and $H || (1, 1, 1)$, heat capacity $C_p$  measured to 9 T are always consistent with $M/H$ (Fig.~\ref{fig:F2}). The antiferromagnetic phase boundary persists to 14 T with $T_N$ changing by $\sim 0.5$ K (Fig.~\ref{fig:F2}). Both heat capacity and magnetic susceptibility indicate that the magnetic phase transition is always second order and there is no additional phase transition over the whole $H$-$T$ space. For all three crystalline directions, the phase diagrams $T_N(H)$ is isotropic, indicating isotropic (Heisenberg type) spin interactions in ZnFe$_2$O$_4$. $H$-$T$ phase diagrams along all three directions support a single-$Q$ antiferromagnetic state. 

Spins separated by $(a, a, a)$ would be of opposite sign for all three vectors of the triple-$Q$ state. The magnetic cell is thus a primitive but not body-centered type for spins. Using \textit{JANA}2020, we refine for both double-$Q$ and triple-$Q$ spin structures with various irreps of several allowed magnetic space groups such as $P_c\bar{4}21m$, $P_c\bar{4}2m$, $P_I\bar{4}3m$, $P_I\bar{4}3n$, and $P_I 213$. Both $R$ and $GOF$ values do not improve with the additional degrees of freedom, and these refinements all generate moments ranging from zero to full sizes on independent spin sites. We conclude a multi-$Q$ state is not favored.

\subsection{Stacked ferromagnetic and antiferromagnetic tetrahedra}

All of the possible refined magnetic structures of ZnFe$_2$O$_4$, in the  different magnetic space groups discussed above (Fig. \ref{fig:F6} and  Table~\ref{app:Table2}), are non-collinear and of three-dimensional type. We note here the common features of these spin structures, as they represent the definitive characteristics of antiferromagnetism in ZnFe$_2$O$_4$. First, spins lie predominantly within the $a$-$b$ plane, as the $c$-axis component in all kernel/epikernel models varies only from zero to ~0.5 $\mu_B$. Second, all refined spin structures can be viewed as consisting of eight spin clusters, each with four nearly parallel (ferromagnetically aligned) spins (Fig. \ref{fig:F6}); these 8 FM tetrahedra are arranged antiferromagnetically within the magnetic unit cell.

In Fig. \ref{fig:F4}a, the measured intensities of lattice and magnetic reflections are plotted against $(h^2+k^2+l^2)$ or equivalently $q_{hkl}^2$, with $q_{hkl}$ the wave vector of reflection $(h, k, l)$. While magnetic diffraction intensities $I_M$ evolve by the overall $q$-dependent magnetic form factor $I_M\sim e^{-q^2}$, they also demonstrate significant variations as a function of $(h, k, l)$,  reflective of the magnetic structure factor of the order. The magnetic intensities $I_M$ (Fig. \ref{fig:F4}b) can be modeled by a simple expression:
\begin{equation}
    I_M\sim [1 + \cos(\pi h/2)\cos(\pi l/2)],
\label{eq:Im}
\end{equation}
where $h$, $k$, and $l$ are the even, odd, and half-integer indices respectively of the reflection $(h, k, l)$. This form was included in a complex expression for the diffraction intensity in Ref. \cite{Konig1970}. It is possible to track the origin of Eq. \ref{eq:Im} to four spins on a single tetrahedron of the pyrochlore lattice. While  multiple models for the spin structure of individual tetrahedra can generate Eq. \ref{eq:Im}, one scenario that is consistent with all our refinement results is four parallel spins along \textit{any direction}, so ${I_M\sim [\sum_{i=1}^{4}\exp{(i\mathbf{q}_{hkl} \mathbf{r}_i})]^2}$. As the simple expression of Eq. \ref{eq:Im} captures the main features of our experimental results (Fig. \ref{fig:F4}b), the remaining differences are indicative of fine details of antiferromagnetically aligned components. 

Alternatively, the same spin structures can be viewed as a stack of tetrahedra where each is composed of four antiferromagnetically arranged spins in the $a$-$b$ plane (Fig. \ref{fig:F6}). In some spin models, there exists a small ferromagnetic component along the $c$-axis in one type of antiferromagnetic tetrahedra while the other type has a net zero moment; this issue wll be addressed in detail below. We thus have two equivalent perspectives of the spin structure in ZnFe$_2$O$_4$, regarding it as either a set of antiferromagnetically arranged FM spin clusters, or as stacked zero-moment, locally antiferromagnetic tetrahedra. These two perspectives are not contradictory, as the pyrochlore lattice is made of corner sharing tetrahedra positioned in a diamond/zincblende structured network. The nearest neighbor tetrahedra of an AF type are always of FM types, and vice versa, forming a three-dimensional checkerboard pattern. Each individual spin is  situated at the junction of two tetrahedra and is ferromagnetically aligned to three spins on one side, and antiferromagnetically aligned to another three spins on the other side. 

\subsection{Symmetry and choice of the spin structure} 
In order to differentiate between the possibilities and identify the proper magnetic space group and spin structure,  we next explore the evolution of the checkerboard spin structure in ZnFe$_2$O$_4$ as imposed symmetry constraints are reduced.  The four space groups in Fig. \ref{fig:F6} provide at least three layers of symmetry hierarchy. The epikernel groups $I_c \bar{4}2m$ and $I_c \bar{4}2d$ both possess the mirror planes, and between them, they exchange the degrees of freedom of two independent spins to the displacement vector $(a/2, 0, a/2)$ of the parent FCC lattice and one spin. The kernel group $I_c \bar{4}$ loses the mirror planes but keeps the four-fold axis. The lowest symmetry group $I_c 222$ lacks both mirror planes and the four-fold axis. 

Refining in the symmorphic group $I_c \bar{4}2m$, the degeneracy of symmetry operations has spins on the mirror planes confined to directions perpendicular to the planes they are located. Half of the Fe spins in the unit cell would therefore align strictly along either $(1, 1, 0)$ or $(1, -1, 0)$ directions within the $a$-$b$ plane with no $c$-axis component (Table~\ref{app:Table2}), blue spins in Figs. \ref{fig:F6}c and \ref{fig:F6}d). In space group $I_c \bar{4}2m$, the other half of spins (grey ones in Figs. \ref{fig:F6}c and \ref{fig:F6}d) are not located on a mirror plane, and mirror operations restrict them to be parallel to the planes, again along either $(1, 1, 0)$ or $(1, -1, 0)$ directions and also the $c$-axis. This mirror symmetry restriction persists through all irreps with different origins of group $I_c \bar{4}2m$, and is the main reason spins reside in the $a$-$b$ plane and become parallel on a tetrahedron. The dominant components in the $a$-$b$ plane and FM clusters in all refinement models in Fig. \ref{fig:F6}, despite mirror operations are no longer being required in both magnetic space groups $I_c \bar{4}$ and $I_c 222$, can be regarded as a confirmation of the mirror symmetry’s presence in the spin structure. 

In all three space groups $I_c \bar{4}2m$, $I_c \bar{4}$, and $I_c 222$, we notice that the two independent sets of spins each form an antiferromagnetic tetrahedron, which are separated by the displacement vector $(a/2, 0, a/2)$ of the parent FCC lattice (color vs. grey in Fig. \ref{fig:F6}). These two antiferromagnetic spin clusters are of nearly orthogonal relationship as one has spins pointing toward the center of the tetrahedron while the other has spins aligned tangential to the tetrahedron (Fig. \ref{fig:F6}). This degree of freedom of two independent spins provides the possibility for spins to remain aligned closely to the $(1, 1, 0)$ family of directions in both groups $I_c \bar{4}$ and $I_c 222$. In contrast, when the number of independent spins is reduced to one in space group $I_c \bar{4}2d$, the spin alignment is dramatically changed to be mostly along $(1, 0, 0)$ family of directions, despite the presence of mirror operations. The resulting antiferromagnetic tetrahedron is similar to the average of two orthogonal antiferromagnetic configurations in all groups $I_c \bar{4}2m$, $I_c \bar{4}$, and $I_c 222$. Two key components of the spin structure in ZnFe$_2$O$_4$ are thus the mirror symmetry and two independent spin sites. They lead to $(1, 1, 0)$-aligned spins and their configuration can only be destroyed by excessive symmetry constraints during refinement. We can thus identify $I_c\bar{4}2m$ as the sole suitable magnetic space group. 

There remain two distinctive spin structures based on two irreps in  $I_c \bar{4}2m$. Their difference lies in how each independent set of spins form the antiferromagnetic tetrahedra. In Fig. 6, we note that irrep $mW1$ for $I_c \bar{4}2d$, $I_c \bar{4}2m$, and $I_c \bar{4}$ always provides a distorted \textit{all-in-all-out} type of antiferromagnetic configuration (Figs. \ref{fig:F6}a, c, e). By contrast, irrep $mW2$ of all three groups always leads to a \textit{two-in-two-out} configuration (Figs. \ref{fig:F6}b, f). As spin moments are mostly in the $a$-$b$ plane, the \textit{all-in-all-out} and \textit{two-in-two-out} configurations in two dimensions are both antiferromagnetic. However, when spins can be three-dimensional, as in ZnFe$_2$O$_4$ with a small $c$-axis component, the \textit{two-in-two-out} type would introduce a minute yet finite FM moment for each tetrahedron, which eventually averages out across the magnetic unit cell. The \textit{all-in-all-out} type would always sum to a zero net-moment along all directions in all models.

For irreps such as $mW2$, which lead to a \textit{two-in-two-out} type of antiferromagnetism in all three space groups, we notice individual $c$-axis components, at $\sim0 \mu_B$ and $0.40-0.55\mu_B$, are largely different, and are significantly larger than that of \textit{all-in-all-out} spin configuration ($0-0.15\mu_B$) based on irreps $mW1$ and its equivalents ( Table~\ref{app:Table2}). In space group $I_c 222$, where these four irreps no longer apply, the \textit{all-in-all-out} antiferromagnetic configuration persists with a zero net moment, yet the largest $c$-axis component in our refinements rises in the antiferromagnetic configuration of spins tangential to the tetrahedron (Table~\ref{app:Table2}, grey spins of Fig. \ref{fig:F6}g). Spin components along the $c$-axis that are permitted by symmetry would exist because of Dzyaloshinsky’s principle \cite{Dzyaloshinsky1958}, yet large $c$-axis components seem like mathematical artifacts. Irrep $mW1$ in the magnetic space group $I_c \bar{4}2m$ (Fig. \ref{fig:F6}c) thus leads to the physically most sensible spin structure for the antiferromagnetism in ZnFe$_2$O$_4$. We note the irrep choice of $mW1$ or $mW2$ does not affect the discussion from here on.

\begin{figure}[tb]
    \centering
    \includegraphics[width=3in]{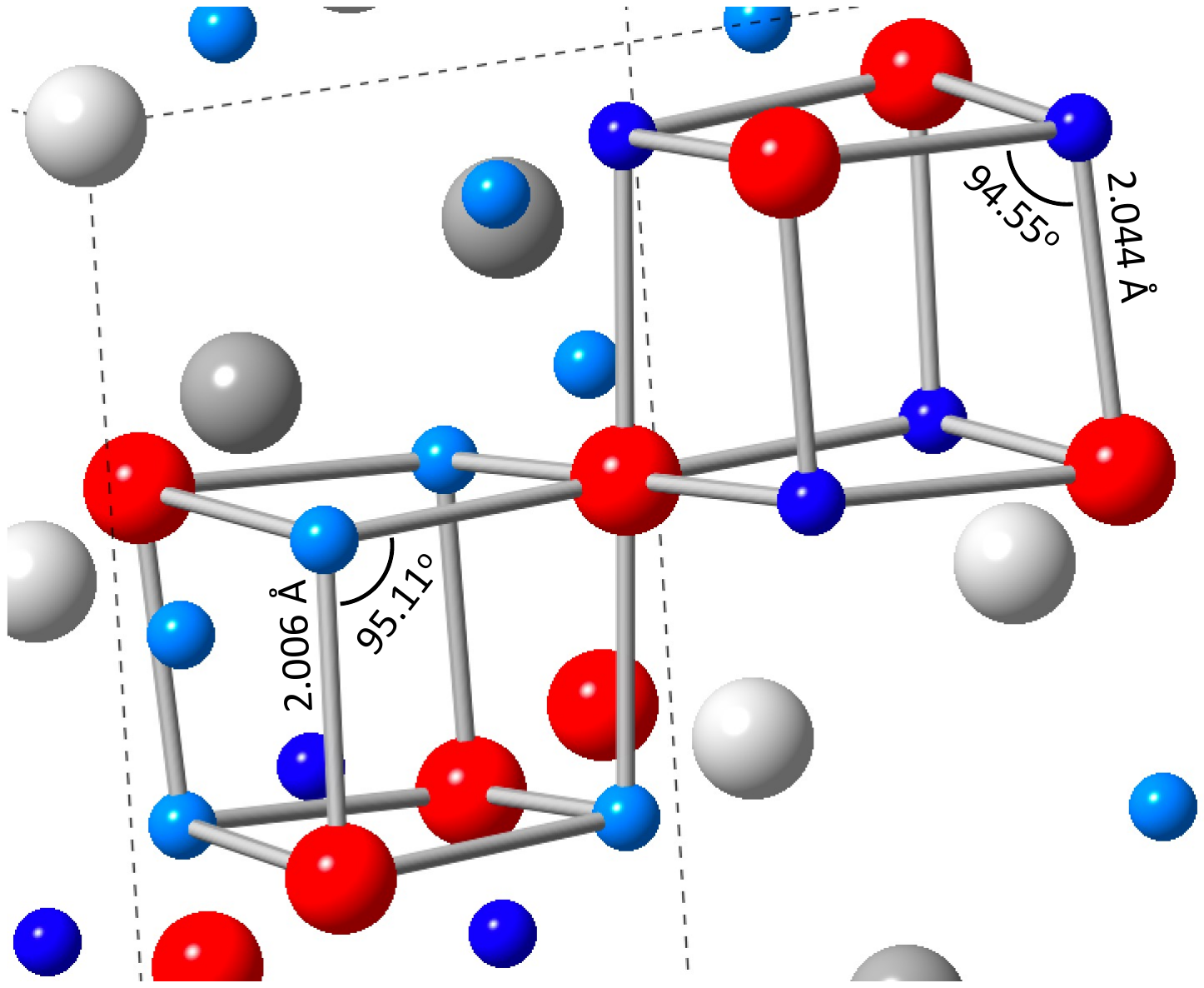}
    \caption{Detailed view of two neighboring Fe tetrahedra in the breathing pyrochlore lattice, based on lattice parameters at 6 K as listed in Table~\ref{app:Table1} The atomic species are: Fe (red), O1 (light blue), O2 (deep blue), Zn1 (light grey) and Zn2 (deep grey). Here the Fe-O bond lengths differ by 2\% between large and small tetrahedra, while the Fe-O-Fe bond angle is larger by $\sim0.6^\circ$ in the smaller tetrahedra. The straight O1-Fe-O2 bond angles are $180^\circ$ to a resolution of $0.01^\circ$. }
    \label{fig:F7}
\end{figure}

\subsection{Breathing lattice and the checkerboard pattern}

The phenomenon of two contrasting types of tetrahedra with only one species of spins is not often observed in long-range ordered pyrochlore magnets \cite{Aroyo2006}. For Ising-type all-in-all-out antiferromagnets such as Cd$_2$Os$_2$O$_7$ \cite{Yamaura2012}, the spin structure on tetrahedra is uniform in the bulk, varying at ferromagnetic domain walls which would not be counted as a stable thermodynamic phase \cite{Wang2020b,Feng2021}. With a Curie-Weiss temperature $T_{CW}\sim-25$ K and $T_{N}=9.95$ K, the spin structure in ZnFe$_2$O$_4$ is not expected to be greatly frustrated. Instead, the viability of coexisting  FM and AF tetrahedra stems from a degree of freedom that is already present in the lattice: the space group $F \bar{4}3m$ of broken inversion symmetry. 

Because of the breathing mode of the pyrochlore lattice, every spin is located at the joint point of one large and one small tetrahedron. Its coupling to neighbor spins depends on both  Fe-O-Fe super-exchange and direct Fe-Fe exchange, in a delicate balance among many different orbit choices \cite{Dronova2022}. The lattice structure based on CORELLI data refinement (Table~\ref{app:Table1}) is drawn in Fig. \ref{fig:F7}. The Fe-O and Fe-Fe distances both differ by ~2\% between two different-sized tetrahedra, suggesting stronger exchange couplings for spins on smaller tetrahedra. Furthermore, the Fe-O-Fe bond angle $95.11^\circ$ on the small tetrahedra is about 0.6$^\circ$ larger than the equivalent bond angle on the large tetrahedra. In our previous refinement \cite{Dronova2022}, the relationship between these two angles is opposite with the small tetrahedra hosting a narrower Fe-O-Fe angle. As bond angles are sensitively dependent on details of refined atomic positions, here our improved, full refinement clarifies this relationship. The Fe-O-Fe angle at $\sim 95^\circ$ is close to the boundary between ferromagnetic and antiferromagnetic types of super-exchange interaction \cite{Dronova2022}. For parallel spins on the large tetrahedra, the exchange interaction would be weak even when it is ferromagnetic, and the increasing Fe-O-Fe angle on small tetrahedra would support stronger antiferromagnetic interaction \cite{Dronova2022}. Both the negative $T_{CW}$ and the long-range ordered ground state indicate the antiferromagnetic interaction should be the type that dominates.

\section{Discussion}

The refined staggered moment of the antiferromagnetic order is relatively small, averaging only $\sim3.0 \mu_B$ per Fe ion (Table~\ref{app:Table2}). This is consistent with the moment sizes reported by previous neutron magnetic powder diffraction studies \cite{Fayek1970,Konig1970,Boucher1970}. Our previous study of the Curie-Weiss behavior \cite{Dronova2022}  revealed an individual moment of 5.53 $\mu_B$ per Fe ion, very close to the expected 5.92 $\mu_B$ of $s=5/2$. In all previous neutron magnetic scattering experiment with strongly disordered specimens\cite{Hastings1953, Fayek1970, Boucher1970, Konig1970, Schiessl1996, Burghart2000, Usa2004, Kamazawa1999, Kamazawa2003, Dronova2022, Sandemann2022}, a large amount of magnetic diffuse scattering was observed. Accordingly, some of the missing moments might be recovered in the elastic channel as frozen (static) short-range correlations. For our clean-limit single crystals with no observable magnetic diffuse scattering (Fig. \ref{fig:F1}), the missing moments would exist in the inelastic (dynamic) channel, with some as long-range coherent magnons \cite{Chan2023}. 

\begin{figure}[!tb]
    \centering
    \includegraphics[width=3.5in]{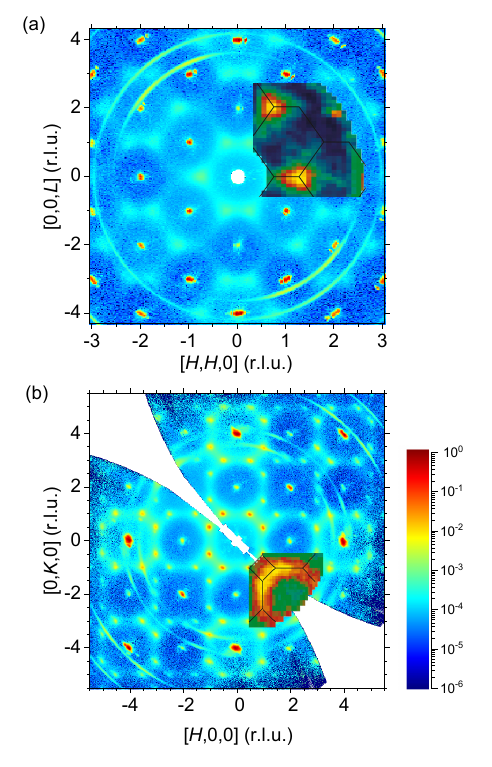}
    \caption{A comparison of the elastic magnetic diffuse scattering of ZnFe$_2$O$_4$ (full frames) at 6 K and inelastic spin scattering of ZnCr$_2$O$_4$ (partial overlays) at $\sim$1meV and 15 K. (a) A view of the reciprocal zone spanned by the $(H, H, 0)$ and $(0, 0, L)$ axes. (b) A view of the reciprocal zone spanned by the $(H, 0, 0)$ and $(0, K, 0)$ axes. The magnetic diffuse scattering was measured from a specimen with a high level of inversion disorder so the spectral weight of static spin correlation spreads more visibly across  reciprocal space, and has little difference below and above $T_N$.  The diffuse data in panel (b) was previous reported in Ref. \cite{Dronova2022} and rescaled for the plots here; the data in panel (a) is not previously published. Intensities are scaled relative to the highest-count single pixel. Our magnetic diffuse scattering data from disordered crystals \cite{Dronova2022} is similar to results reported in Ref. \cite{Kamazawa2003}. Both sets of ZnCr$_2$O$_4$ inelastic scattering data are adapted from Ref. \cite{Lee2002}}
    \label{fig:F8}
\end{figure}

In the classical spin liquid systems ZnCr$_2$O$_4$ and MgCr$_2$O$_4$, dispersionless flat bands of excitations were observed surrounding the $(2, 2, 0)$ reciprocal space point \cite{Lee2000, Lee2002, Tomiyasu2013, Gao2018}. This special feature of inelastic scattering was attributed to either a local mode based on a hexagonal ring of spins named hexamer \cite{Lee2002,Gao2018}, or several local zero modes of hexamer and heptamer \cite{Tomiyasu2013,Gao2018}, in addition to spin waves based on multiple interactions between spins on the six-spin ring \cite{Bai2019}. However, in both ZnCr$_2$O$_4$ and MgCr$_2$O$_4$, the magnetic ground state is complicated, with multiple $k$-orders such as $(1/2, 1/2, 0)$, $(0, 1, 1/2)$, and $(0, 0, 1)$ \cite{Gao2018, Bai2019}. In comparison, ZnFe$_2$O$_4$ has a cubic lattice and a single magnetic structure, which allows us to offer a simple alternative explanation for the excitation in chromates based on the antiferromagnetic ground state of $(1, 0, 1/2)$ order.

In Eq. (1) of Ref. \cite{Lee2002}, an analytical expression was used to describe the inelastic spectral weight $|F_6 (Q)|^2$, with the index 6 indicating the hexagonal ring structure. Ref. \cite{Lee2002} claimed that a hexamer is necessary as this spectral form cannot originate from a four-spin structure on one octahedron. However, we note that $|F_6 (Q)|^2$  can be simplified to $|F_6(Q)|^2\sim 2[1-\cos(\pi h/2)\cos(\pi l/2)]$ with $h$, $k$, $l$ following the even, odd, and half integer notation in our Eq. \ref{eq:Im}. In such a fashion, $|F_6 (Q)|^2$ is directly related to our elastic spin scattering intensity $I_M$ in Eq. \ref{eq:Im} as $2I_M=4-|F_6 (Q)|^2$; here we take unspecified coefficients of both $I_M$ and $|F_6 (Q)|^2$ as unity, and neglect the $q$-dependent spin form factor. A delta function of spin correlations in real space $\langle S_\mathbf{r} S_\mathbf{r^\prime}\rangle = \delta(\mathbf{r}-\mathbf{r^\prime})$, i.e. a fully disordered, uncorrelated spin state, would Fourier transform into a uniform spin density in  reciprocal space, $\langle S_q^2\rangle = \mathrm{const}$. As spin moments are summed over both static and dynamic channels, the inelastic spectrum of ZnCr$_2$O$_4$ should be complementary to a $(1, 0, 1/2)$ ordered antiferromagnetic structure such as that of ZnFe$_2$O$_4$. As the mathematical structure of $I_M$ can be derived from four parallel spins on one tetrahedron, the inelastic spectrum in Ref. \cite{Lee2002} could have a simple explanation of uncondensed fluctuations in the paramagnet that compensate a four-spin FM cluster. 

This comparison of analytical forms of $I_M$ in Eq. \ref{eq:Im} and $|F_6 (Q)|^2$ in Ref. \cite{Lee2002} can be directly visualized in the spectral distributions of the spin density in  reciprocal space (Fig. \ref{fig:F8}). The inelastic patterns of ZnCr$_2$O$_4$ in Figs. 3a, b of Ref. \cite{Lee2002} are fully complementary to the spin elastic diffuse scattering pattern of a ZnFe$_2$O$_4$ specimen with a high level of disorder \cite{Dronova2022} (Fig. \ref{fig:F8}). The static spin spectral weight centers around $(4, 0, 0)$ and $(4, 4, 0)$ positions (Figs. \ref{fig:F1}, \ref{fig:F8}). The inelastic scattering in Fig. \ref{fig:F8}b can be viewed as eight equivalent $(1, 0, 1/2)$ $W$ points surrounding the $(2, 2, 0)$ position and the smearing of spectral weight could be due to short range spin correlations originated from structural/chemical disorder. Our explanation of the excitation spectrum should also apply to MgCr$_2$O$_4$, as higher energy excitation bands \cite{Gao2018} have much weaker spectral weight than that of the first resonance band. In Ref. \cite{Bai2019}, a spin wave spectrum was built on a Hamiltonian utilizing four interactions over three neighbor distances, which can be similarly applied to ZnFe$_2$O$_4$. The structure of MgCr$_2$O$_4$ does not have a breathing lattice to accommodate the $(1, 0, 1/2)$ type of spin order as the ground state, so a large structural distortion and multiple magnetic wave vectors emerge below $T_N$. While both Refs. \cite{Gao2018, Bai2019} recognize $(1, 0, 1/2)$ as part of the puzzle, our results suggest this order should be the major magnetic instability of all these chromate and ferrite systems and can potentially explain most features of their inelastic spectra. Inelastic resonance bands in those chromates have been attributed to local zero spin modes such as the six-spin ring mode, leading to assignments of those systems as classical spin liquids. A full exploration of fluctuations complementing to the long-range, $(1, 0, 1/2)$ vector modulated, non-collinear spin structure in ZnFe$_2$O$_4$ could shed lights into these molecular local modes and classical spin liquids. It remains to be verified whether inelastic neutron scattering of ZnFe$_2$O$_4$ crystals of minimal disorder would be instrument resolution-limited or dispersionless such as those in ZnCr$_2$O$_4$ and MgCr$_2$O$_4$. Such a study can potentially bring a comprehensive understanding of the role of chemical disorder.

\begin{acknowledgements}
We thank Yishu Wang for insightful discussions, and C. M. Hoffmann for assistance of data collection at TOPAZ. Y. Feng acknowledges financial support from the Okinawa Institute of Science and Technology Graduate University, with subsidy funding from the Cabinet Office, Government of Japan. A portion of this research used resources at the Spallation Neutron Source, a US Department of Energy Office of Science User Facility operated by the Oak Ridge National Laboratory. Development of JANA is continuously supported by the Czech Science Foundation and Academy of Sciences of the Czech Republic. The work at Caltech was supported by AFOSR, grant FA9550-20-1-0263.

\end{acknowledgements}

\appendix

\section{\label{app:reduce}Neutron diffraction data reduction}
The typical process of data reduction for time-of-flight (TOF) neutron diffraction is illustrated in Ref. \cite{Sullivan2018}. The detector electronics  at both beamlines allow each neutron’s wavelength to be recorded based on its arrival time at the detector. The information of both wavelength and the detector’s position allows all diffraction events to be converted into  reciprocal space without knowledge of the $UB$ matrix. For measurements at TOPAZ, it is possible to integrate regions of dense counts before determining the primitive (Niggli) cell and in turn the orientational $UB$ matrix. For CORELLI, a preliminary $UB$ matrix is typically established before the extended measurement in order to verify the correct zone of interest. The current data reduction script provides a final optimization of the $UB$ matrix for datasets from both beamlines \cite{Morgan}. The finalized $UB$ matrix provides the measured lattice constants and their uncertainties.  

\begin{table*}[!tb]
    \centering
    \caption{Lattice structure refined by \textit{JANA}2020 in the space group  $F\bar{4}3m$. Zn ions occupy sites Zn1:(0, 0, 0) and Zn2:(0.25, 0.25, 0.25) with symmetry of $4a$ and $4c$ respectively. The lattice structure measured at CORELLI was refined first before the magnetic refinement. Measurements at both beamlines used the same piece of single crystal.\label{app:Table1}}
    \begin{tabular}{c|c|c}
Beamline of Measurement&TOPAZ&CORELLI\\
\midrule
Measurement Temperature (K)& 100 & 6\\
Wavelength range (Å)&	0.6-3.&	0.8-2.4\\
d-spacing (Å) range&	$\geq$0.5	&$\geq$0.7\\
Signal/uncertainty ratio $I$/$\delta I$&	$\geq$3&	$\geq$3\\
Reflection events&	5199	&9511\\
Independent reflections&	2759	&859\\
Reflection families&	144	&59\\
Lattice constant $a$ (Å)&	8.4383(2)	&8.4337(6)\\
Extinction type&	Secondary, mixed type, Lorentzian mosaic&	Secondary, mixed type, Lorentzian mosaic\\
Atomic position (16e) of iron&	0.62457(9)&	0.62408(9)\\
1st atomic position (16e) of oxygen&	0.38625(9)&	0.38678(9)\\
2nd atomic position (16e) of oxygen&	0.86529(9)&	0.86610(9)\\
$U_{iso}$(Fe)&	0.00155(5)&	0.00047(9)\\
$U_{iso}$(O1)&	0.00283(18)&	0.00156(23)\\
$U_{iso}$(O2)&	0.00242(18)&	0.00181(24)\\
$U_{iso}$(Zn1)&	0.00221(23)&	0.00025(12)\\
$U_{iso}$(Zn2)&	0.00150(20)&	0.00025(12)\\
$R$&	4.18&	5.84\\
Goodness of Fit&	2.15&	2.58\\
\bottomrule
\end{tabular}
    \label{tab:T1}
\end{table*}

The conversion to  reciprocal space collapses all single crystal diffraction events of one specific reflection into one single volume \cite{Sullivan2018}. After that, a proper treatment of events with a spread of neutron wavelengths is the central issue. If the integration range is overly narrow, the listed events would have a poor signal versus background ratio. On the other hand, if the integration is over a wide wavelength range, diffraction signals with largely different extinction effects are summed together, making the potential correction inaccurate. For our monolithic single crystal, the mosaic across the whole sample volume is of  order of 0.02$^\circ$ to 0.05$^\circ$ FWHM, much smaller than the neutron beam divergence which is of  order of 5$^\circ$ FHWM. Thus, the incident beam’s angular divergence determines the total spread of neutrons wavelengths as a diffraction “event” for one reflection at the specific angular position. For TOPAZ, placements of the sample’s angular positions are far apart. For CORELLI, as the rotational step is only 1.5$^\circ$, diffraction events on neighboring frames can be correlated. 

Previously, for time-of-flight neutron diffraction at TOPAZ, each event of reflection $(h, k, l)$ was integrated with one-dimensional profile fitting along the wave vector $Q$ direction \cite{Schultz2014}, using a Gaussian profile convoluted with an exponential form to account for the tail at the long-wavelength side. This integration scheme was applied to our previous analysis in Ref. \cite{Dronova2022}). Here, we improve the integration scheme to profile fitting in the three-dimensional reciprocal space, and the new data reduction procedure now applies to neutron diffraction at both TOPAZ and CORELLI \cite{Morgan}. It is similar to the procedures described in Ref. \cite{Sullivan2018} but with a few variations. The 3-D profile fitting first is carried out for all events of one reflection. The data set summed over all events allows the best statistics to determine the peak profile. This is essential for CORELLI as it employs Helium-3 tubes as detectors; these have lower efficiency compared to to the scintillator-based Anger cameras at TOPAZ. While Ref. \cite{Sullivan2018} would fit the 2D detector’s transverse directions with bivariate Gaussian forms in addition to the time-of-flight profile, here we define the three axes of the ellipsoid fully in the reciprocal space by $Q_P$, $Q_1$, and $Q_2$. The primary wavevector $Q_P$ is the radial vector of the $(h, k, l)$ reflection, while $Q_1$ and $Q_2$ are vectors transverse (perpendicular) to $Q_P$. 

Once the profile envelope is established, one can divide different wavelength contribution to a given reflection into subsets of the whole. For both TOPAZ and CORELLI, a natural division is the individual event of one frame at a fixed sample angular position; combining diffraction events of the same reflection on two neighboring frames at CORELLI introduces further complications for samples with strong extinction. To integrate individual subset, the width parameters of the full profile fitting are used, in order to avoid large fluctuations of the integrated Q-volume associated with reduced counting statistics. 

After further corrections such as the detector efficiency and the Lorentz factor for TOF type of neutron diffraction, the diffraction data is reduced to line entries tabulated with the $h, k, l$ indices of the reflection, integrated intensity $I$ and its measurement uncertainty $\delta I$, central neutron wavelength $\lambda$ of the event, $2\theta$ angle, together with other detector-related information \cite{Morgan}. A single reflection $(h, k, l)$ can have several independent events recorded with different central wavelengths. For cubic systems, there exist many equivalent reflections of permutated $h, k, l$ values, so one family of equivalent reflections could have tens of events across a wide wavelength range (Fig. \ref{fig:F3}). Integrated intensities from two beamlines are very consistent (Fig. \ref{fig:F3}), despite  utilizing different types of neutron detectors and the data having very different statistics because of the measurement time and pattern.

\begin{table*}[!tb]
    \centering
    \caption{\label{app:Table2}Refinement results of the antiferromagnetic structure. One or two independent spins exist for each magnetic space group at $M_1$ (colored), and $M_2$ (grey) in Fig. 6. × indicates zero based on the symmetry forbidden condition. The refined parameters for $mW1$ and $mW2$ are listed here, as $mW3$ and $mW4$ are respectively degenerate with equivalent spin structures and similar parameters.}
    \begin{tabular}{c|c|c|c|c|c|c|c}
Beamline&\multicolumn{7}{|c}{CORELLI}  \\
\midrule
$T$ (K)&\multicolumn{7}{|c}{6}  \\
$\lambda$ range (\r{A})&\multicolumn{7}{|c}{0.8-2.4}  \\
Magnetic reflection events&\multicolumn{7}{|c}{7921}  \\
Independent reflections&\multicolumn{7}{|c}{899}  \\
Magnetic reflection families&\multicolumn{7}{|c}{36}  \\
\midrule
Magnetic space group& $I_c\bar{4}2d$ & $I_c\bar{4}2d$ & $I_c\bar{4}2m$ & $I_c\bar{4}2m$ & $I_c\bar{4}$ &$I_c\bar{4}$ &$I_c222$\\
Representations & $mW1$ \& $mW4$ & $mW2$ \& $mW3$ & $mW1$ \& $mW4$ & $mW2$ \& $mW3$ & $mW1$ \& $mW4$ & $mW2$ \& $mW3$ & \\
\midrule
$M_{1a} (\mu_B)$& 	3.0663 (4)	&3.0374 (5)	&-2.4065 (8)	&1.8263 (8)	&-1.913 (16)	&2.04 (2)	&2.441 (3)\\
$M_{1b} (\mu_B)$&	-0.3960 (12)&	0.4163 (11)&$-M_{2a}$ &$-M_{2a}$ & 2.836 (12)	&-1.63 (2)	&0.753 (7)\\
$M_{1c} (\mu_B)$&	-0.0882 (15)&	-0.4022 (14)&	×&	×	&0.146 (16)	&-0.03 (14)	&0.022 (2)\\
Total $M_1 (\mu_B)$&	3.0930 (19)	&3.0920 (18)	&3.4033 (12)	&2.5827 (11)	&3.42 (3)	&2.61 (14)	&2.554 (8)\\

$M_{2a} (\mu_B)$&	×	&×	&1.8595 (9)	&-2.4006 (8)	&-2.115 (15)	&-2.247 (19)	&-3.084 (6)\\
$M_{2b} (\mu_B)$&	×	&×	&$M_{2a}$ 	&$M_{2a}$	&-1.52 (2)	&-2.561 (19)	&1.344 (11)\\
$M_{2c} (\mu_B)$&	×	&×	&-0.1186 (19)	&0.5326 (18)	&0.122 (3)	&0.550 (12)	&0.968 (4)\\
Total $M_2 (\mu_B)$& 	×	&×	&2.632 (12)	&3.437 (2)	&2.61(3)	&3.45 (3)	&3.501 (14)\\
$R_{lat}$	&6.22	&6.23	&5.86	&5.85	&5.86	&5.85	&5.86\\
$R_{mag}$	&8.79	&8.64	&8.72	&8.55	&8.73	&8.49	&8.73\\
$GOF$	&2.64	&2.59	&2.55	&2.50	&2.55	&2.51	&2.53\\
\bottomrule
    \end{tabular}
    \label{tab:T2}
\end{table*}

\section{\label{app:extinct}Neutron lattice refinement with extinction correction}
The measured diffraction intensity $I_M$ is related to the structure factor $F_{hkl}(\lambda)$ and its theoretically expected value $F_{hkl}$ as $I_M\sim F_{hkl}^2 (\lambda)=y(\lambda)F_{hkl}^2$, with the factor $y(\lambda)$ defined as the extinction factor. This reduction mechanism originates from Darwin’s dynamic diffraction theory for perfect crystals, which leads to the primary extinction $y_p$ \cite{Becker1974}.  For mosaic crystals, the extinction $y_s$ is termed secondary and has a convoluted effect between the limit of mosaic angular spread (type I) and the domination of coherent domain size (type II). In the limit of infinitely-large coherent domains, type II crystals approach perfect crystals. For type I crystals, the difference in the mosaic shape separates this category into Gaussian and Lorentzian types, which should become equivalent in the limit of type II and perfect crystals when the coherent domain size dominates. Ref. \cite{Becker1974} fits their numerically calculated extinction factors to a general analytical form:
\begin{equation}
y_{p,s}=\left[1+2x\frac{A(\theta) x^2}{1+B(\theta)x)}\right]^{-1/2},
\label{eq:extinct}
\end{equation}
as $A(\theta)$, $B(\theta)$, and $x$ have different expressions for different extinction types (primary, and secondary of types I, mixed, and II) and different mosaic shapes (Gaussian vs. Lorentzian). Being a reflection-dependent variable, $x$ includes the structure factor $F_{hkl}$, wavelength $\lambda$, diffraction angle $2\theta$, the coherent particle size $R$, the mosaic width $g$, and other parameters such as polarization $K$ and unit cell volume $V$. The term $2x$ in the bracket can be slightly modified to $2.12x$ for crystals of Gaussian shaped mosaics.

Despite many theoretical studies \cite{Lawrence1973, Becker1974, Tomiyoshi1980, Delapalme1988, Schultz2014}, experimental characterizations of the extinction effect remain limited. Many studies can only correct for the extinction effect with $y>0.30$ in Ref. \cite{Lawrence1973, Delapalme1988}, $y>0.40$ in Ref. \cite{Jauch1988}, and $y>0.50$ in Ref. \cite{Schultz2014}. Many diffraction experiments also utilize only a few wavelengths over a limited range; for example, three neutron diffraction experiments using 0.527, 0.757, and 1.05 \text{\AA} wavelengths were combined to analyze the extinction effect in Ref. \cite{Delapalme1988}. Over such a small range of $y$ and $\lambda$, theoretical models cannot be well differentiated and predict $F_{hkl}^2$ in the $\lambda=0$ limit. 

In Ref. \cite{Dronova2022}, the analytical expression of Eq. \ref{eq:extinct} was applied to each individual family to extract $F_{hkl}^2$ in the $\lambda=0$ limit. However, there are two scenarios in which the evaluation of the extinction effect by Eq. \ref{eq:extinct} becomes difficult. About 25\% of the strong reflection families with a large extinction effect cannot be properly portrayed by Eq. \ref{eq:extinct}, even though they are measured over a large range of wavelength. For reflections of large transferred momenta, diffraction events can only be measured over a very narrow range of $\lambda$. Thus, the fitting of Eq. \ref{eq:extinct} produces parameters with large uncertainties. For the first scenario, it is customary to remove certain families of strong reflections from the refinement \cite{Sequeira1972}. For the second scenario, fitting each reflection family with Eq. \ref{eq:extinct} to extract $F_{hkl}^2$ can only be performed when $y(\lambda)$ is determined by fitting Eq. \ref{eq:extinct} to other medium intensity reflection families measured over a large $\lambda$ range \cite{Dronova2022}. We seek a global refinement with a variable extinction effect $y(\lambda,F_{hkl}^2)$ incorporated in the software. We also intend to take all diffraction events into consideration. 

There are several publicly accessible refinement packages that can treat an input data set with a broad range of wavelengths, such as \textit{Shelxl}, \textit{GSAS}, and \textit{JANA}. For extinction correction, we note \textit{SHELXL} (2014 version) uses a simplified version of Eq. \ref{eq:extinct} without the term containing $A(\theta)$ and $B(\theta)$, and only the expression of $x$ for type I crystal with Gaussian type of mosaic. The term with $A(\theta)$ and $B(\theta)$ is to capture the extinction behavior at large $x$ values, and for mosaic dominated crystals, no large extinction effect is expected. So both features in \textit{SHELXL} were designed to handle only weak extinction effects. For \textit{GSAS} (2004 manual), the listed expressions of $x$ for TOF neutron diffraction are questionable, as TOF type of measurements should not have different $x$ than that of continuous-wave neutron diffraction \cite{Tomiyoshi1980}. Here, we work with the software package \textit{JANA}2020 (Version 1.3.51, Aug. 11th, 2023) \cite{Petricek2023}. In early versions of the software, the extinction parameter $rho_{iso}$ was defined as the ratio of coherent domain size to wavelength $\lambda$. In the current version, wavelength $\lambda$ is separated from $rho_{iso}$ in order to properly handle a data set with continuous wavelength; $rho_{iso}$ now represents only the coherent domain size. The results of the refinement for the lattice are given in Table~\ref{app:Table1} and for the magnetic structure in Table~\ref{app:Table2}. 

The broken inversion symmetry of the $F\bar{4}3m$ space group introduces two structural domains. Domain degeneracy in \textit{JANA} is typically handled as a transpose matrix for indices. However, Friedel’s law nearly always holds for neutron nuclear diffraction \cite{Petricek2010}, as the neutron atomic form factor is a constant independent of $q$. This is different from the x-ray form factor, which has an anomalous behavior near absorption edges so the imaginary part of the charge form factor $f^{\prime\prime}$ can break Friedel’s law to make the diffraction intensities of $(h, k, l)$ and $(-h, -k, -l)$ different for systems with broken inversion symmetry. Here we refine the single crystal diffraction data under the $F\bar{4}3m$ space group as one single domain. 

Current extinction correction in \textit{JANA}2020 utilizes the analytical form of Eq. \ref{eq:extinct} while adapting to different scenarios. While Eq. \ref{eq:extinct} is an analytical expression and intuitive, it is only an approximation of numerical calculations in Ref. \cite{Becker1974}. For example, the analytical forms of $B(\theta)$ for the Lorentzian type of mosaic \cite{Becker1974} has a gap at $2\theta=90^\circ$, and further discontinuities in its derivative of $\theta$ at $2\theta=90^\circ$. To improve the modeling of extinction correction, it is desirable to directly apply the tabulated numerical results of $y$ in Ref. \cite{Becker1974} in future. This simplifies the choice scenarios to only secondary extinctions of Gaussian and Lorentzian mosaic crystals with the lowest $y$ values down to 0.025, and 0.09 respectively (Tables 3, 4 of Ref. \cite{Becker1974}).

%

\end{document}